\documentclass[10pt, conference, compsocconf]{IEEEtran}
\usepackage[table,xcdraw]{xcolor}	
\usepackage{todonotes}  
\usepackage{amsmath}    
\usepackage[vlined, ruled, boxed]{algorithm2e}
\SetAlFnt{\small}		
\SetAlCapFnt{\small}
\SetAlCapNameFnt{\small}
\usepackage{multirow}	
\usepackage[tight,footnotesize]{subfigure}
\usepackage{fancyhdr}			
\usepackage[hidelinks]{hyperref} 			

\makeatletter
\def\ps@IEEEtitlepagestyle{
  \def\@oddfoot{\mycopyrightnotice}
  \def\@evenfoot{}
}
\def\mycopyrightnotice{
  {\footnotesize
  \begin{minipage}{\textwidth}
  \centering
 ~\copyright~2018 IEEE. Personal use of this material is permitted. Permission from IEEE must be  obtained for all other uses, in any current or future media, \\ including reprinting/republishing this material for advertising or promotional purposes, creating new  collective works, for resale  \\ or redistribution to servers or lists, or reuse of any copyrighted component of this work in other works.
 
 DOI: 10.1109/GIOTS.2018.8534539
\end{minipage}
  }
}

\begin{document}
%
\title{Analytic Analysis of Narrowband IoT Coverage Enhancement Approaches}


\author{\IEEEauthorblockN{Pilar Andres-Maldonado\IEEEauthorrefmark{1}\IEEEauthorrefmark{2}, Pablo Ameigeiras\IEEEauthorrefmark{1}\IEEEauthorrefmark{2}, Jonathan Prados-Garzon\IEEEauthorrefmark{1}\IEEEauthorrefmark{2}, Juan J. Ramos-Munoz\IEEEauthorrefmark{1}\IEEEauthorrefmark{2}, \\ Jorge Navarro-Ortiz\IEEEauthorrefmark{1}\IEEEauthorrefmark{2}, and Juan M. Lopez-Soler\IEEEauthorrefmark{1}\IEEEauthorrefmark{2}}
\IEEEauthorblockA{\IEEEauthorrefmark{1}Research Center on Information and Communication Technologies \\
\IEEEauthorrefmark{2}Department of Signal Theory, Telematics and Communications \\
University of Granada\\
Granada, Spain\\
Email:  \{pilaram, pameigeiras, jpg, jjramos, jorgenavarro, juanma\}@ugr.es \IEEEauthorrefmark{1}\IEEEauthorrefmark{2}}
}


\maketitle

\begin{abstract}
The introduction of Narrowband Internet of Things (NB-IoT) as a cellular IoT technology aims to support massive Machine-Type Communications applications. These applications are characterized by massive connections from a large number of low-complexity and low-power devices. One of the goals of NB-IoT is to improve coverage extension beyond existing cellular technologies. In order to do that, NB-IoT introduces transmission repetitions and different bandwidth allocation configurations in uplink. These new transmission approaches yield many transmission options in uplink. In this paper, we propose analytical expressions that describe the influence of these new approaches in the transmission. Our analysis is based on the Shannon theorem. The transmission is studied in terms of the required Signal to Noise Ratio, bandwidth utilization, and energy per transmitted bit. Additionally, we propose an uplink link adaptation algorithm that contemplates these new transmission approaches. The conducted evaluation summarizes the influence of these approaches. Furthermore, we present the resulting uplink link adaptation from our proposed algorithm sweeping the device's coverage.
\end{abstract}


%
\IEEEpeerreviewmaketitle

\section{Introduction}
\label{sec:intro}


The Internet of Things (IoT) concept embodies the vision of everything connected. This vision encompasses a vast ecosystem of emerging use cases in markets such as industrial machinery, healthcare, smart cities, etc. Generally, IoT use cases can be divided into two categories: massive Machine-Type Communications (mMTC), and Ultra-Reliable and Low Latency MTC (URLLC). mMTC is characterized by massive connections from a large number of low-complexity and low-power devices. Conversely, URLLC requires high reliability and support for extreme latency requirements. Focusing on mMTC, its requirements comprise a great challenge on current existing mobile networks due to their high cost and high power consumption. To solve that, the Third Generation Partnership Project (3GPP) started a feasibility study on providing cellular IoT connectivity. The aim was to find a solution competitive in the Low Power Wide Area (LPWA) segment. Among the solutions proposed, Narrowband IoT (NB-IoT) technology emerged in Release 13. NB-IoT is based on Long Term Evolution (LTE) specification and reuses several technical components. NB-IoT uses a new radio design that enables a wide range of IoT applications in the licensed spectrum that are low cost, use low data rates, require long battery life and often operate in remote and deep indoor areas. To satisfy these characteristics, the main goals of NB-IoT are \cite{3gpp45820}:
\begin{itemize}
\item Low device cost: NB-IoT removes many features of LTE to be kept as simple as possible.
\item Improved indoor coverage: +20 dB compared to legacy General Packet Radio Service (GPRS), corresponding to a Maximum Coupling Loss (MCL) of 164 dB.
\item Low power consumption: 10 years of battery life with a battery capacity of 5 Wh.
\item Massive connections: 52547 connections per cell site sector. Additionally, NB-IoT can be easily scaled up by adding more carriers as capacity requires.
\end{itemize}

Particularly, to achieve the significant coverage enhancement in uplink, NB-IoT uses transmission repetitions and different network's bandwidth allocation configurations. These new transmission approaches yield an intricate system with many transmission options, where it is necessary to balance the trade-off between the transmission reliability and throughput of the network. For this goal, this paper proposes analytical expressions that describe the impact of the new transmission approaches of NB-IoT. The analysis is based on the Shannon theorem and is done for three transmission properties: required Signal to Noise Ratio (SNR), bandwidth utilization, and energy per transmitted bit. Additionally, we propose a new link adaptation algorithm for NB-IoT considering these new transmission approaches. The presented results summarize the influence of these approaches for the transmission properties analyzed. Furthermore, we present the resulting uplink link adaptation from our proposed algorithm sweeping the device's coverage.

The remainder of the paper is organized as follows. Section \ref{sec:background} gives an introduction to NB-IoT and related works. Section \ref{sec:sys_model} describes the system model. Section \ref{sec:math_model} includes the analytical analysis. Section \ref{sec:LinkAdapt} introduces the proposed link adaptation algorithm. Section \ref{sec:results} presents the numerical results. Finally, Section \ref{sec:conclusion} sums up the conclusions.

\section{Background and related works}
\label{sec:background}

The new NB-IoT radio interface uses a dedicated carrier of 180 kHz that can be deployed in-band of LTE, in a guard band or stand alone. In Release 13 and 14, NB-IoT only supports Frequency Division Duplex (FDD) half-duplex mode. In downlink, Orthogonal Frequency-Division Multiple Access (OFDMA) is applied using a 15 kHz subcarrier spacing with 14 symbols used to span a subframe of 1 ms. In uplink, Single-Carrier Frequency-Division Multiple Access (SC-FDMA) is applied, using either 3.75 kHz or 15 kHz subcarrier spacing. NB-IoT supports both single-tone and multi-tone operations in uplink. Particularly for multi-tone uplink transmission (12, 6 or 3 tones), only 15 kHz subcarrier spacing is allowed. The smallest unit in uplink to map a transport block is a Resource Unit (RU). The definition of an RU depends on the Narrowband Physical Uplink Shared CHannel (NPUSCH) format and subcarrier spacing. Furthermore, the duration of the RU depends on the tones allocated. For more information, see \cite{3GPP36211}.

\subsection{Coverage enhancement approaches}
\label{subsec:cov_approach}

Targeting a significant coverage improvement in NB-IoT is achieved by signal repetitions, new control channels, and specifically for uplink, bandwidth reduction. Figure 1 shows an example of the possible configuration of a User Equipment (UE) transmission to enhance its coverage. Specifically for the NPUSCH format 1 responsible for uplink data transmission, the repetitions have two possible Redundancy Versions (RV). The arrangement of the repetitions depends on the number of tones, subcarrier spacing, and number of repetitions \cite{WPNB}. 
In addition to repetitions, NB-IoT enables a set of possible allocations of the network bandwidth. Single-tone configurations are mandatory and provide capacity in signal-strength-limited scenarios. Multi-tone configurations are optional and provide higher data rates for UEs in good coverage. Note that both approaches entail an increase of the time needed to finish the transmission. 

Under the targeted low range of SNR, an accurate channel estimation becomes a dominant issue that limits the coverage improvement \cite{3gpp151216}. In these radio conditions, the performance of the channel estimator is expected to be poor. For more information, see \cite{Beyene17}.

\begin{figure}[!bt] 
	\centering
    \includegraphics[width=0.73\columnwidth]{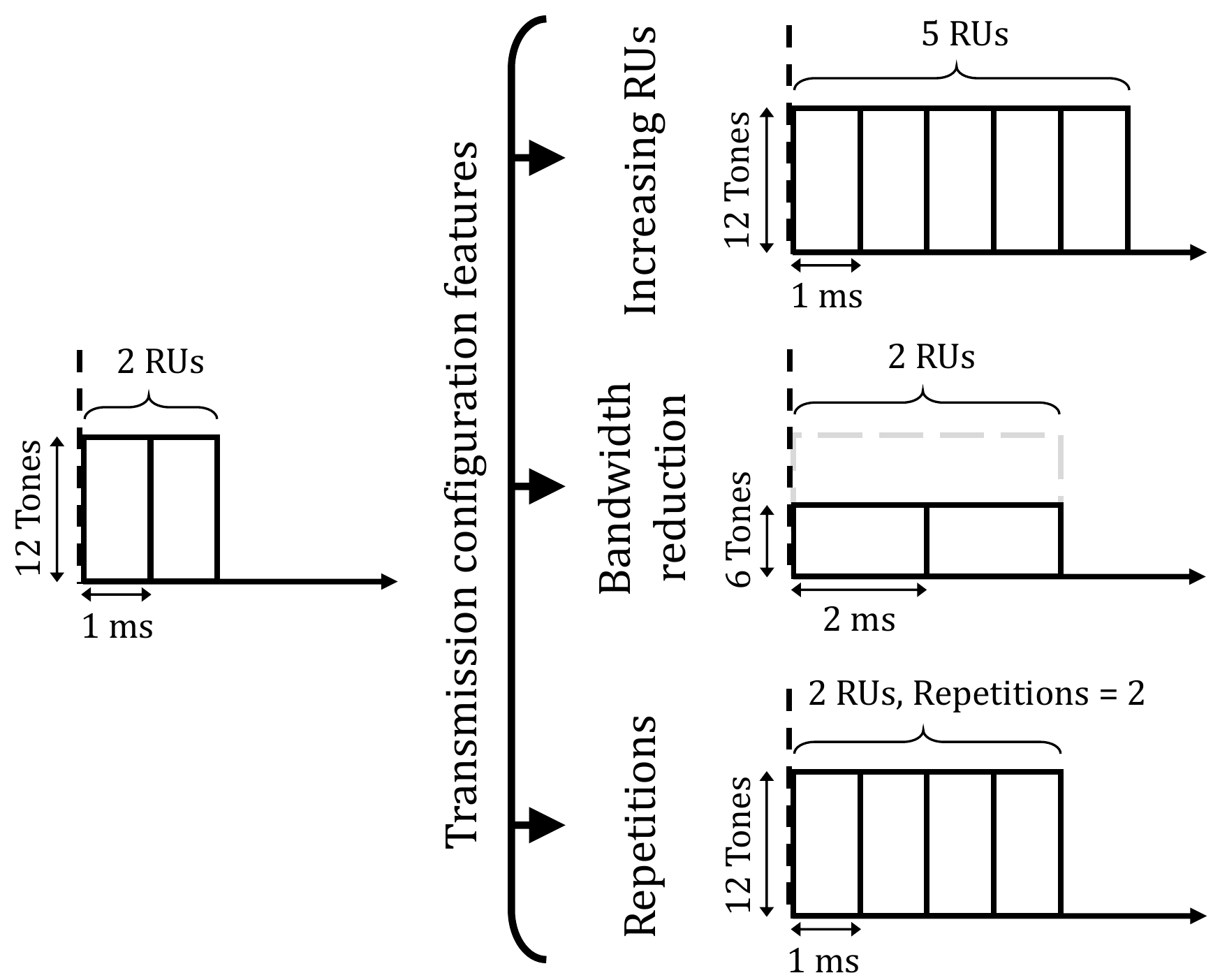}
	\caption{Example of NPUSCH transmission approaches in NB-IoT.}
	\label{fig:radioUL}
\end{figure}

\subsection{Related works}
\label{subsec:rel_works}

As a newly-developing technology, NB-IoT has still several issues to investigate. On the one hand, regarding NB-IoT performance, the authors of \cite{Adhikary16} evaluate the coverage performance of NB-IoT. In \cite{Lauridsen16}, the authors also included results of capacity for a small rural area with real operator-deployed base stations. Nevertheless, these works and other similar provide final results for specific configurations that hinder the comparison of the results. Particularly, the authors of \cite{Beyene17} derive and simulate an analytical bound for the SNR gain from repetitions. On the other hand, the extensive use of repetitions in NB-IoT adds a new dimension for link adaptation that traditional LTE mechanisms do not consider. The works \cite{Mu16} and \cite{Yu17} examine this topic. The former proposed a downlink link adaptation algorithm. The latter investigated dynamic uplink link adaptation. However, neither of them examine analytically the impact of the repetitions. Additionally, the uplink link adaptation proposed in \cite{Yu17} do not contemplate the bandwidth reduction as a new dimension in uplink link adaptation.

In our previous work \cite{AndresNET17}, we analyzed the energy consumption of the data transmission procedures for NB-IoT UEs in three specific levels of coverage. We also provided an analysis of the radio resources utilization in different transmission scenarios. In this work, we delve into the configuration of the NB-IoT UE's transmission and the influence of the new features of NB-IoT. Furthermore, we analyze a large range of coverage conditions.

\section{System model}
\label{sec:sys_model}

\begin{figure}[!bt] 
	\centering
    \includegraphics[width=0.85\columnwidth]{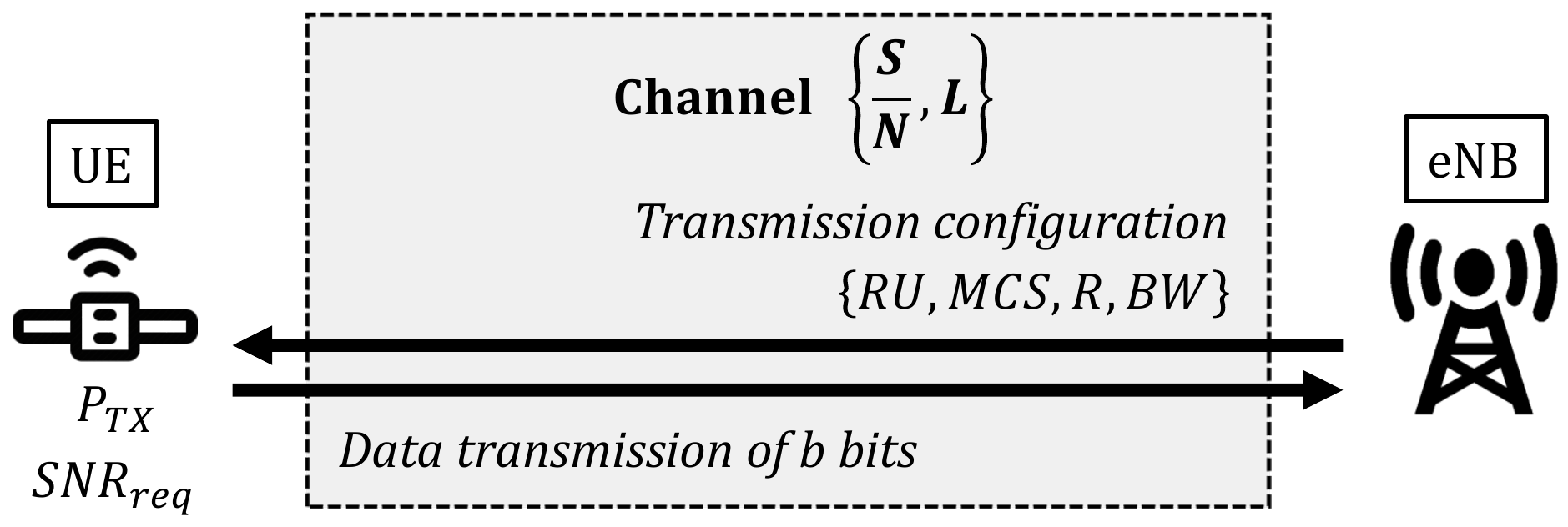}
	\caption{System model.}
	\label{fig:sysModelLink}
\end{figure}

Let us consider a transmission of a single packet of size $b$ bits from an NB-IoT UE to its eNB. We assume a very slowly time-variant channel with losses due to path loss, denoted as $L$. As we only consider channel losses because of $L$, then $MCL = L$. To compensate channel losses, the UE adjusts its transmission power $P_{tx}$ up to a maximum allowed value $P_{max}$. Therefore, the signal to noise ratio of the channel, denoted as $\frac{S}{N}$, can be calculated as

\begin{equation}
\frac{S}{N} = \frac{P_{tx}}{L \cdot F \cdot N_{o} \cdot BW }
\label{eq:SNRchannel}
\end{equation}

where $N_{o}$ is the thermal noise density, $F$ is the receiver noise figure, $BW = SCS \cdot N_{T}$ is the allocated bandwidth, $SCS$ is the subcarrier spacing, and $N_{T}$ is the number of tones. The overall system assumed is depicted in Figure \ref{fig:sysModelLink}.

When the eNB configures the UE's data transmission, we analyze three independent approaches of the transmission. In the first case, the eNB only modifies the number of RUs allocated to the UE. In the second case, the eNB simply reduces the bandwidth allocated. Finally, in the third case, the eNB only modifies the number of repetitions assigned. When applying repetitions, we assume all repetitions have the same RV. Therefore, the same information is repeated in each repetition and combined at the eNB using chase combining. For all cases, along with the number of RU, there is a corresponding Modulation and Coding Scheme (MCS) level. In this work, we only consider QPSK modulation. The MCS and number of RUs define the code rate of the transmission $CR$, obtained as 

\begin{equation}
  CR = \frac{b + CRC}{RU \cdot \frac{symbols}{RU}  \cdot \frac{bits}{symbol}}
  \label{eq:CR}
\end{equation}

where $CRC$ is the size in bits of the Cyclic Redundancy Check code, and $RU$ is the number of RUs allocated to the UE. NB-IoT allows a range of possible values of $RU$. Likewise, the combination of the MCS, number of RUs, and allocated bandwidth determine the data rate of the transmission derived as

\begin{equation}
  R_{b} = \frac{b + CRC}{RU \cdot T}
  \label{eq:bitrate}
\end{equation}

where $R_{b}$ is measured in bits per second, and $T$ is the duration in seconds of an RU. Note that the duration of the RU depends on the bandwidth allocated to the UE. As the number of tones decreases, $T$ increases. Herein, we denote this dependency on the bandwidth as $T^{(BW)}$. 

The resulting configuration of the transmission parameters determines the required SNR of the UE's transmission, denoted as $SNR_{req}$. In order to decode UE's uplink transmission successfully, the $SNR_{req}$ is bounded by $SNR_{req} \leq \frac{S}{N}$. When applying repetitions or bandwidth reduction, the $SNR_{req}$ can be reduced. For simplicity, we assume ideal channel estimation. Therefore, there is an ideal gain in the $SNR_{req}$ when both approaches are used.

\section{Analytical model}
\label{sec:math_model}

This section focuses on the impact of the multiple parameters in the uplink transmission. To do that, our analytical model is based on the Shannon theorem. The analysis of this section is separated into three independent transmission approaches: i) RU number modification; ii) bandwidth reduction; and iii) repetitions. These three approaches are represented in the analysis as $(.)^{(RU,BW,R)}$, respectively. Finally, the last subsection joins the three previous analysis and studies the transmission power required to achieve this reasoning. 

\subsection{RU number modification}
\label{subsec:RU_analysis}

Let first study the influence of the RUs in the transmission. To ease the analysis, we assume the bandwidth is fixed to its maximum allowed value $BW_{max}$, and the repetitions to its minimum $R = 1$. Let $SNR_{req}$ be the required SNR of the UE's transmission. Using the Shannon bound, $SNR_{req}$ can be written as

\begin{equation}
  \begin{aligned}
    SNR_{req}^{(RU,BW_{max},1)} &= 2^{R_{b}^{(RU,BW_{max},1)}/BW_{max}} - 1 \\
    &= 2^{\frac{b + CRC}{BW_{max} \cdot RU \cdot T^{(BW_{max})}}} - 1 
  \end{aligned}
  \label{eq:SNRreq_wRU}
\end{equation}

From the data rate of the UE, we can obtain the bandwidth utilization $\gamma$ of the transmission. Consequently,

\begin{equation}
  \begin{aligned}
    \gamma^{(RU,BW_{max},1)} &= \frac{R_{b}^{(RU,BW_{max},1)}}{BW_{max}} \\
    &= \frac{b + CRC}{BW_{max} \cdot RU \cdot T^{(BW_{max})}}
  \end{aligned}
  \label{eq:gamma_wRU}
\end{equation}

Furthermore, let $\frac{E_{b}}{N_{o}} = \frac{SNR_{req}}{\gamma}$ be the lower bound of the received energy per bit to noise power spectral density ratio, and $\hat{E_{b}}$ be the energy per transmitted bit, then

\begin{equation}
  \begin{aligned}
    \hat{E_{b}}^{(RU,BW_{max},1)} &= \frac{E_{b}}{N_{o}}^{(RU,BW_{max},1)} \cdot L \cdot F \cdot N_{o} 
    \\
    &= \frac{2^{\frac{b + CRC}{BW_{max} \cdot RU \cdot T^{(BW_{max})}}} - 1 }{\frac{b + CRC}{BW_{max} \cdot RU \cdot T^{(BW_{max})}}} \cdot L \cdot F \cdot N_{o}
  \end{aligned}
  \label{eq:EbNo_wRU}
\end{equation}

Observe that as the number of RUs is greater, the rest of the transmission properties analyzed (i.e. $SNR_{req}$, $\gamma$, and $\hat{E_{b}}$) decrease.

\subsection{Bandwidth reduction}
\label{subsec:BW_analysis}

Let us now study the influence of the bandwidth reduction in the transmission. Again, we assume the rest of approaches of the transmission have fixed values. Then, the number of RUs and repetitions equal to their minimum, $RU = 1$ and $R = 1$, respectively. Repeating the previous analysis, we start with the required SNR expressed in this case as

\begin{equation}
  \begin{aligned}
    SNR_{req}^{(1,BW,1)} &= 2^{R_{b}^{(1,BW,1)}/BW} - 1 \\
    &= 2^{\frac{b + CRC}{BW \cdot T^{(BW)}}} - 1 
  \end{aligned}
  \label{eq:SNRreq_wBW}
\end{equation}

The bandwidth utilization is obtained as follows:

\begin{equation}
  \begin{aligned}
    \gamma^{(1,BW,1)} &= \frac{R_{b}^{(1,BW,1)}}{BW} = \frac{b + CRC}{BW \cdot T^{(BW)}}
  \end{aligned}
  \label{eq:gamma_wBW}
\end{equation}

Now, the energy per transmitted bit can be derived as

\begin{equation}
  \begin{aligned}
    \hat{E_{b}}^{(1,BW,1)} &= \frac{E_{b}}{N_{o}}^{(1,BW,1)} \cdot L \cdot F \cdot N_{o} 
    \\
    &= \frac{2^{\frac{b + CRC}{BW \cdot T^{(BW)}}} - 1 }{\frac{b + CRC}{BW \cdot T^{(BW)}}} \cdot L \cdot F \cdot N_{o}
  \end{aligned}
  \label{eq:Etb_wBW}
\end{equation}

Note that for all multi-tone configurations, the increase of the RU's duration $T^{(BW)}$ is the same as the reduction in the bandwidth. Therefore, while the UE maintains a multi-tone configuration, the analyzed transmission properties preserve their values. This holds true if the transmission power is reduced in accordance with the bandwidth. However, when moving from multi-tone to single-tone configuration, the increase of $T^{(BW)}$ and the reduction of the bandwidth is unequal. Thereby, single-tone configurations present higher $SNR_{req}$, $\gamma$, and $\hat{E_{b}}$ than multi-tone configurations. If the transmission power is maintained, this approach concentrates the limited power on a narrower bandwidth. This enhances the received SNR, and thus the coverage can be extended.

\subsection{Repetitions}
\label{subsec:Rep_analysis}

Finally, let us now study the impact of the repetitions in the transmission. When applying repetitions, the received transmission's copies at the eNB can be combined. Therefore, the resulting SNR after the combining, denoted as $SNR_{c}$, is the sum of the SNRs of each repetition received at the eNB. Consequently,

\begin{equation}
  \begin{aligned}
    SNR_{c} = \sum^{R} SNR_{req} = R \cdot SNR_{req}
  \end{aligned}
  \label{eq:SNRc_wR}
\end{equation}

From (\ref{eq:SNRc_wR}) and the fact that $SNR_{c}$ is equal to the Shannon bound, we can derive the required SNR as

\begin{equation}
  \begin{aligned}
    SNR_{req}^{(1,BW_{max},R)} &= \frac{2^{R_{b}^{(1,BW_{max},1)}/BW_{max}} - 1}{R} \\
    &= \frac{ 2^{\frac{b + CRC}{BW_{max} \cdot T^{(BW_{max})}}} - 1}{R}
  \end{aligned}
  \label{eq:SNRreq_wR}
\end{equation}

where $(.)^{(1,BW_{max},R)}$ denotes the number of RUs is equal to its minimum $RU = 1$, and the bandwidth is equal to its maximum allowed value $BW_{max}$. Again, the bandwidth utilization $\gamma$ including repetitions can be expressed as

\begin{equation}
  \begin{aligned}
    \gamma^{(1,BW_{max},R)} &= \frac{R_{b}^{(1,BW_{max},1)}}{R \cdot BW_{max}} = \frac{b + CRC}{R \cdot BW_{max} \cdot T^{(BW_{max})}}
  \end{aligned}
  \label{eq:gamma_wR}
\end{equation}

Finally, the energy per transmitted bit can be derived as

\begin{equation}
  \begin{aligned}
    \hat{E_{b}}^{(1,BW_{max},R)} &= \frac{E_{b}}{N_{o}}^{(1,BW_{max},R)} \cdot L \cdot F \cdot N_{o} 
    \\
    &= \frac{2^{\frac{b + CRC}{BW_{max} \cdot T^{(BW_{max})}}} - 1}{\frac{b + CRC}{BW_{max} \cdot T^{(BW_{max})}}} \cdot L \cdot F \cdot N_{o}
  \end{aligned}
  \label{eq:Etb_wR}
\end{equation}

Note that $\hat{E_{b}}^{(1,BW_{max},R)}$ is no longer a function of the number of repetitions. The utilization of repetitions reduces the $SNR_{req}$ at the expense of the reduction of $\gamma$.

\subsection{Combined analysis}
\label{subsec:comb_analysis}

In practice, the eNB will configure the transmission considering all approaches explained in previous subsections. For that reason, in this subsection, we generalize the previous analytical expressions taken into account all approaches. To begin with, the required SNR can be expressed as

\begin{equation}
  \begin{aligned}
    SNR_{req}^{(RU,BW,R)} &= \frac{2^{R_{b}^{(RU,BW,1)}/BW} - 1}{R} \\
    &= \frac{ 2^{\frac{b + CRC}{BW \cdot RU \cdot T^{(BW)}}} - 1}{R}
  \end{aligned}
  \label{eq:SNRreq_wALL}
\end{equation}

The bandwidth utilization is as follows:

\begin{equation}
  \begin{aligned}
    \gamma^{(RU,BW,R)} &= \frac{R_{b}^{(RU,BW,1)}}{R \cdot BW} = \frac{b + CRC}{R \cdot BW \cdot RU \cdot T^{(BW)}}
  \end{aligned}
  \label{eq:gamma_wALL}
\end{equation}

Additinally, the energy per transmitter bit is calculated as

\begin{equation}
  \begin{aligned}
    \hat{E_{b}}^{(RU,BW,R)} &= \frac{E_{b}}{N_{o}}^{(RU,BW,R)} \cdot L \cdot F \cdot N_{o} 
    \\
    &= \frac{2^{\frac{b + CRC}{BW \cdot RU \cdot T^{(BW)}}} - 1}{\frac{b + CRC}{BW \cdot RU \cdot T^{(BW)}}} \cdot L \cdot F \cdot N_{o}
  \end{aligned}
  \label{eq:Etb_wALL}
\end{equation}

From (\ref{eq:SNRreq_wALL}), (\ref{eq:gamma_wALL}), and (\ref{eq:Etb_wALL}), we can observe the impact of each approach in the analyzed transmission properties.
Concluding our analysis, if we assume $\frac{S}{N} = SNR_{req}^{(RU,BW,R)}$, the transmission power must be: 

\begin{equation}
  \begin{aligned}
    P_{tx} = SNR_{req}^{(RU,BW,R)} \cdot L \cdot F \cdot N_{o} \cdot BW
  \end{aligned}
  \label{eq:Ptx_wALL}
\end{equation}

Thereby, reducing the $SNR_{req}$ or $BW$ implies a reduction of the $P_{tx}$. Note that if the $P_{tx}$ reaches its upper limit (i.e. $P_{tx}= P_{max}$), the path loss is upper bounded as

\begin{equation}
  \label{eq:PLmax_wALL}
  \begin{aligned} 
  L_{max} = \frac{P_{max}}{F \cdot N_{o} \cdot BW \cdot SNR_{req}^{(RU,BW,R)}}
  \end{aligned}
\end{equation}

Since $L_{max}$ is the maximum supported path loss, if the UE experiences a path loss greater than $L_{max}$, the eNB will not be able to decode the transmission correctly. To solve that, the UE will need to extend its coverage by means of reducing its $SNR_{req}$ through repetitions or increasing RUs, or decreasing its allocated bandwidth.

\section{Model application: Link Adaptation}
\label{sec:LinkAdapt}

As an example of application of the previous analysis, we combined it with our proposed link adaptation algorithm. For that purpose, we consider uplink link adaptation can be performed in three dimensions: i) MCS or RUs, ii) bandwidth allocated, and iii) repetitions. In order to perform link adaptation, we assume three phases. Firstly, from the 3GPP's Transport Block Size (TBS) table for NB-IoT \cite{3GPP36213}, we calculate the $R_{b}$ corresponding to each combination of MCS and number of RUs allowed. Secondly, from the analysis of Section \ref{sec:math_model}, we estimate the $SNR_{req}$ of each position of the TBS table. This is done considering bandwidth reduction and repetition approaches too. Later, the algorithm searches the optimal configuration using as a criterion the minimization of the transmission time. The algorithm takes as inputs the needed SNR $SNR_{in}$, and the size of the packet $b$. The following Algorithm \ref{al:linkadaptation} shows the pseudo code of our proposed link adaptation.  
As a result of the three considered dimensions in uplink link adaptation, our algorithm relies on different flags to search more than one possible solution. Then, from the found solutions, the optimal solution is selected. Note that we give priority to bandwidth reduction as it preserves the bandwidth utilization.

\begin{algorithm}[b!]
\SetInd{0.4em}{0.35em}
\LinesNumbered
\label{al:linkadaptation}
\caption{Proposed Uplink Link Adaptation Algorithm for NB-IoT}
\SetKwInOut{Input}{Input}
\SetKwInOut{Output}{Output}
\newcommand\mycommfont[1]{\textit{#1}} 
\SetCommentSty{mycommfont}
\newcommand{\pushline}{\Indp}
\newcommand{\popline}{\Indm}
\Input{$SNR_{in}$ and $b$}
\Output{$Best\_Point\_Found \left ( C, I_{MCS}, I_{RU}, R_{idx} \right )$}
\BlankLine
	$N_{confs} \leftarrow 5$ \tcp{Bandwidth configurations allowed}
    $C \leftarrow 1 $ \tcp{Current bandwidth configuration to evaluate}
    $R_{max} \leftarrow$ Maximum number of repetitions allowed \\
    $RU_{max} \leftarrow 8$ \tcp{Maximum number of RUs}
    $SNR_{req}^{C}(I_{MCS}, I_{RU}, R) \leftarrow SNR_{req}$ of the configuration $C$ corresponding to MCS level $I_{MCS}$, $I_{RU}$ RUs and $R$ repetitions\\  
    $TBS(I_{MCS}, I_{RU}) \leftarrow$ number of bits corresponding to $I_{MCS}$ combined with $I_{RU}$ in NB-IoT TBS table\\
\BlankLine
  \While{$C \leq N_{confs}$  \& not finished}{
    Set $R_{max}$ according to $C$
    
    \If{$length \left ( Points\_Found \right ) > 1$ }{
       Reinitialization of some parameters and flags  
     }

      \While{$R_{idx} \leq R_{max}$ \& not finished}{
      
         \While{$I_{MCS} \leq MCS_{max}$ \& not finished}{
         
             \If{$min(SNR_{req}^{C}(I_{MCS}, :, R_{idx})) \leq SNR_{in}$}{

                \While{$I_{RU} \leq RU_{max}$ \& not finished}{

                   \If{$TBS(I_{MCS},I_{RU}) \geq  b$ \& $SNR_{req}^{C}(I_{MCS},I_{RU}, R_{idx}) \leq SNR_{in}$}{
                      $Points\_Found \leftarrow \left ( C, I_{MCS},I_{RU}, R_{idx} \right )$\\
                      Depending on $C$, some parameters are reconfigured to search more points before increasing $C$ or exit the algorithm
                  }
               }
             }
          }
      }
   }
$Best\_Point\_Found \leftarrow get \left ( min\_transmission\_time \left ( Points\_Found \right ) \right )$\\   

\end{algorithm}

\section{Numerical results}
\label{sec:results}

Herein, in this section we provide numerical results illustrating our previously stated analytical expressions of the transmission properties (i.e. $SNR_{req}$, $\gamma$, and $\hat{E_{b}}$). Table \ref{table:parametersvalues} summarizes the parameters used in our evaluation.

\begin{table}[b!]
\centering
\caption{Variable definitions}
\label{table:parametersvalues}
\resizebox{\columnwidth}{!}{%
\begin{tabular}{l|c|l}
\hline
\textbf{\textbf{Variable}} & \textbf{\textbf{Value}} & \textbf{Description} \\ \hline
$P_{tx}$ & Variable & Transmission power \\
$P_{max}$ & 23 & Maximum $P_{tx}$ (dBm) \cite{3GPP36213} \\
$L$ & Variable & Path loss (dB) \\
$F$ & 3 & Receiver noise figure (dB) \cite{3gpp45820}\\
$N_{o}$ & -174 & Thermal noise (dBm/Hz) \\
$SCS$ & $\left \{ 15, 3.75\right \}$ & Subcarrier spacing (kHz) \\
$N_T$ & $\left \{ 12,6,3,1\right \}$ & Number of tones \\
$CRC$ & 24 & CRC code length (bits) \\
$RU$ & $\left \{ 1,2,3,4,5,6,8,10\right \}$ & Number of RUs \\
$T^{(BW)}$ & $\left \{ 1,2,4,8,32\right \}$ & RU's duration (ms) \\
$R$ & $\left \{ 1,2,4,8,16,32,128\right \}$ & Number of repetitions \\ \hline
\end{tabular}%
}
\end{table}

\subsection{NB-IoT transmission approaches comparison}

\begin{figure}[b!] 
	\centering
    \includegraphics[width=1\columnwidth]{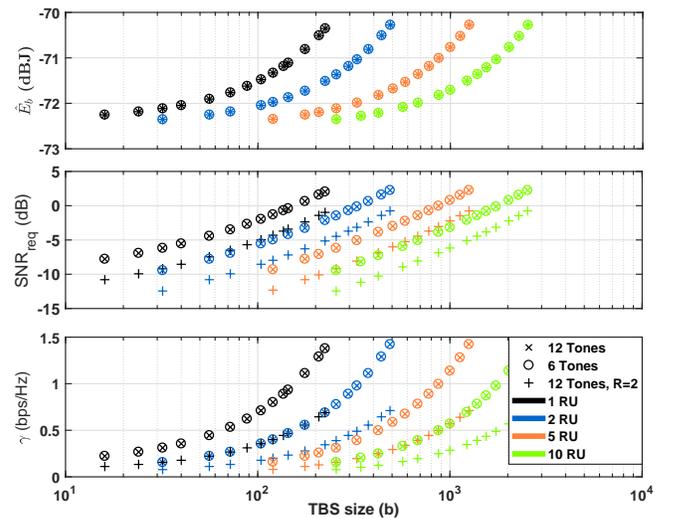}
	\caption{Transmission properties comparison as a function of the TBS size when different transmission approaches are used.}
	\label{fig:basicComparisonv2}
\end{figure}

In the first part of the evaluation, we consider the UE has good coverage. The transmission power follows equation (\ref{eq:Ptx_wALL}) without constraints, and we assume $L=100dB$. Figure \ref{fig:basicComparisonv2} shows the analyzed transmission properties as a function of the TBS size. This figure compares the results when there is a modification in the amount of RUs, a bandwidth reduction, or an increase of repetitions. Due to the power resulting from (\ref{eq:Ptx_wALL}), the utilization of these approaches involves the reduction of the transmission power. For a given TBS size, a greater number of RU achieves a lower energy per transmitted bit $\hat{E_{b}}$. In terms of required SNR, repetitions obtain the best results. However, this is at the expense of reducing the bandwidth utilization. Then, for a UE with good coverage, there is no benefit on applying repetitions.

\subsection{Link adaptation evaluation}

\begin{figure}[tb!]
  \centering
  
  \subfigure[Transmission time.\label{fig:LA_TXtime}]{\includegraphics[width=0.95\columnwidth]{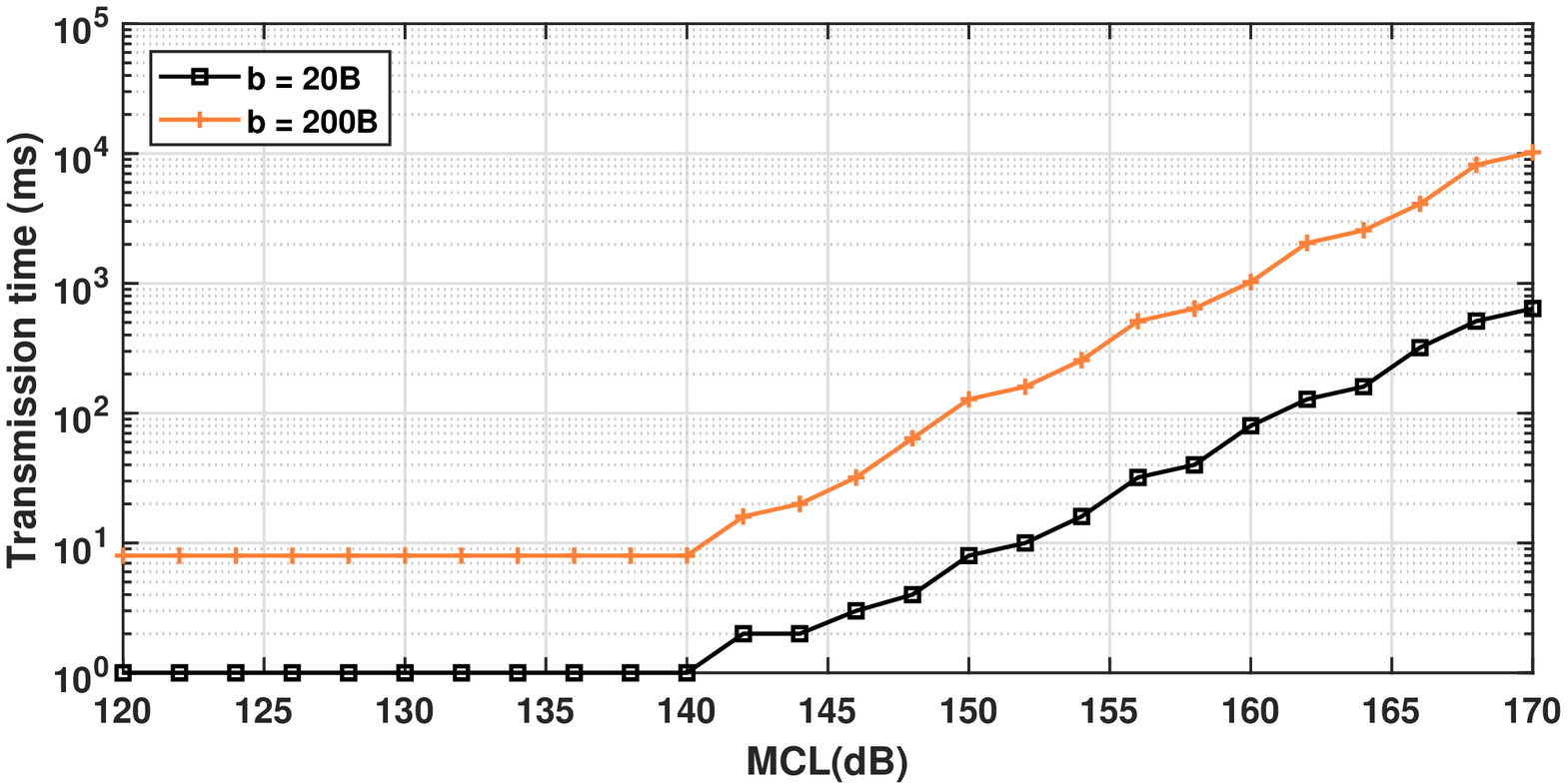}}%
  
  \subfigure[Repetitions and bandwidth allocation.\label{fig:LA_RC}]{\includegraphics[width=1\columnwidth]{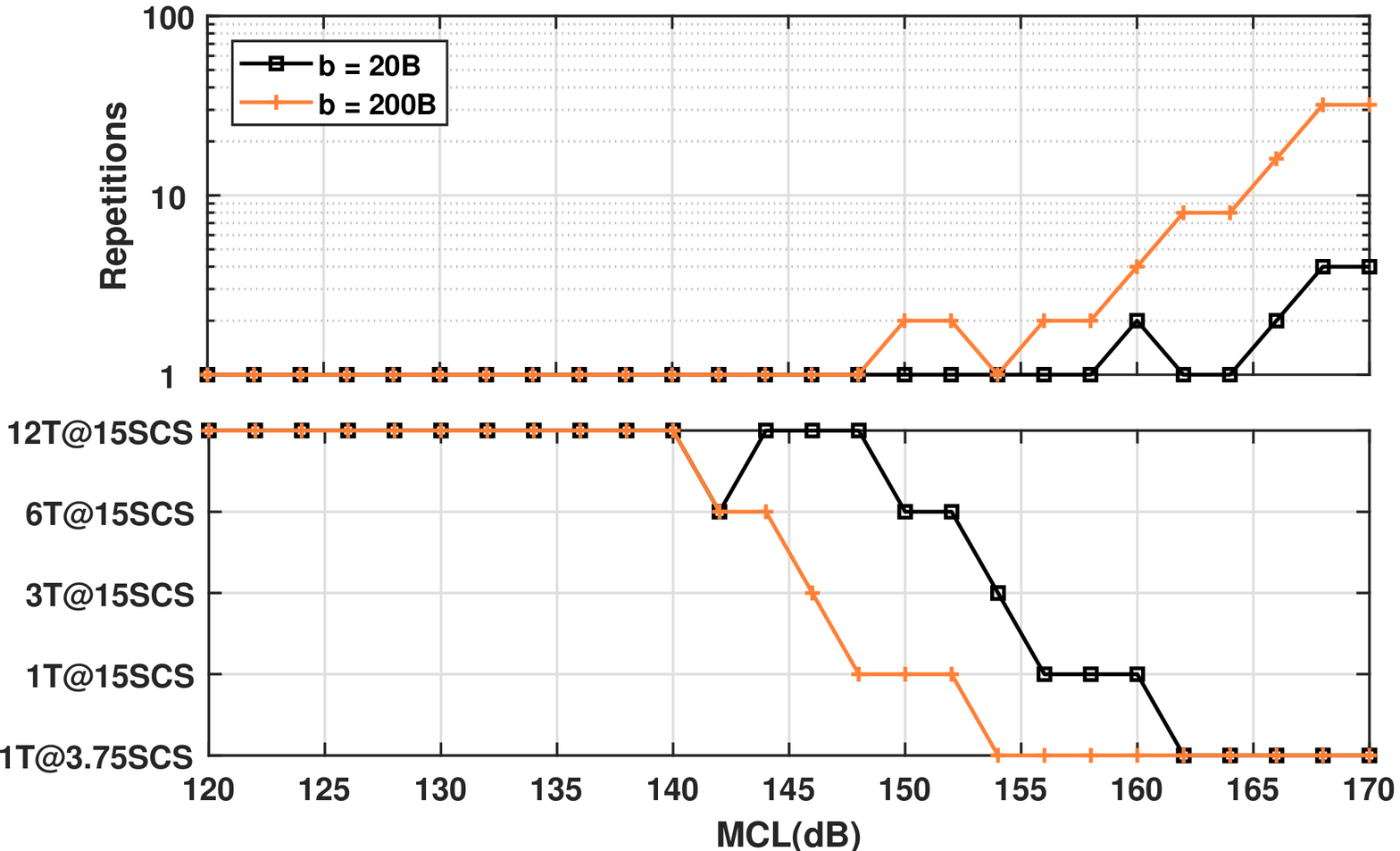}}%
  
  \caption{UE's transmission time \ref{fig:LA_TXtime} and transmission configuration \ref{fig:LA_RC} as a function of the MCL considering two packet sizes  (20B and 200B).}
  \label{fig:LA_ALL}
\end{figure}

We now present the UE's transmission configuration using our proposed link adaptation algorithm and sweeping a range of MCL. In this case, we consider 3GPP's transmission power constraints, such as $P_{max}$. To do that, the UE's NPUSCH transmission power is calculated as stated by the 3GPP \cite{3GPP36213}, assuming $P_{0} = -100dBm$, and $\alpha_{c}=1$. Figure \ref{fig:LA_ALL} depicts the transmission time, repetitions, and bandwidth allocated as a function of the MCL considering two packet sizes. As the UE has worse coverage (i.e. higher MCL), the transmission time increases. This is reasonable since bandwidth reduction and repetitions are used to achieve the coverage extension needed. Owing to bandwidth reduction preserves the bandwidth utilization when applied, the algorithm gives priority to this approach. Therefore, repetitions are not applied until single-tone configurations are used. As expected, different packet sizes present a similar tendency in the link adaptation. Nevertheless, large packet sizes begin to struggle at lower MCLs than small sizes.


\section{Conclusion}
\label{sec:conclusion}

In this paper, we have derived analytical expressions of the required SNR, bandwidth utilization, and energy per transmitted bit considering the new transmission approaches of NB-IoT (i.e. RU number modification, bandwidth reduction, and repetitions). Furthermore, we have proposed a link adaptation algorithm that exploits all transmission approaches and seeks the minimization the transmission time. The conducted evaluation shows the benefits of the transmission approaches when the transmission power is constrained or not. For a UE in good coverage, RUs modification or bandwidth reduction keep the bandwidth utilization and can cover a limited range of required SNR. Additionally, a greater number of RU achieves a lower energy per transmitted bit. However, in the case that a UE exceeds its maximum supported path loss, bandwidth reduction and repetitions become essential to reach greater coverage extension. For future work, we intend to study the coverage extension limitation when considering realistic channel estimation, and its impact on NB-IoT performance.

\section*{Acknowledgment}
This work is partially supported by the Spanish Ministry of Economy and Competitiveness and the European Regional Development Fund (Project TEC2016-76795-C6-4-R), and the Spanish Ministry of Education, Culture and Sport (FPU Grant 13/04833).


\bibliographystyle{IEEEtran}
\bibliography{main}

\end{document}